\documentclass[12pt]{article}

\RequirePackage{etex}

\usepackage[margin=1in]{geometry}  	
\usepackage{amsmath,amsfonts,amssymb,amsthm}
\usepackage{booktabs}
\usepackage{multirow}
\usepackage[onehalfspacing]{setspace} 
\renewcommand{\arraystretch}{0.6} % make sure matrices are not stretched

\usepackage{float}
\usepackage{caption}
\usepackage{graphicx}
\usepackage{adjustbox}
\usepackage{lscape}

\usepackage{adjustbox}

\usepackage{xcolor}

\usepackage{relsize}

\usepackage{bbm}
\usepackage{subfig}

\definecolor{forestgreen}{RGB}{34,139,34}
\usepackage{hyperref}
\hypersetup{
    colorlinks=true,
    linkcolor=magenta,
    filecolor=magenta, 
    citecolor=forestgreen,      
    urlcolor=blue
}

\usepackage{authblk}

\usepackage{amsthm}
\newtheorem{theorem}{Theorem}
\newtheorem{proposition}[theorem]{Proposition}
\newtheorem*{proposition*}{Proposition}

\usepackage{xpatch}
\makeatletter
\xpatchcmd{\proof}{\@addpunct{.}}{\@addpunct{:}}{}{}
\makeatother

\newcommand{\overbar}[1]{\mkern 1.5mu\overline{\mkern-1.5mu#1\mkern-1.5mu}\mkern 1.5mu}

\makeatletter
\newcommand{\vast}{\bBigg@{3}}
\newcommand{\Vast}{\bBigg@{4}}
\makeatother

\makeatletter
\newcommand*{\indep}{%
  \mathbin{%
    \mathpalette{\@indep}{}%
  }%
}
\newcommand*{\nindep}{%
  \mathbin{%                   % The final symbol is a binary math operator
    \mathpalette{\@indep}{\not}% \mathpalette helps for the adaptation
  }%
}
\newcommand*{\@indep}[2]{%
  \sbox0{$#1\perp\m@th$}%        box 0 contains \perp symbol
  \sbox2{$#1=$}%                 box 2 for the height of =
  \sbox4{$#1\vcenter{}$}%        box 4 for the height of the math axis
  \rlap{\copy0}%                 first \perp
  \dimen@=\dimexpr\ht2-\ht4-.2pt\relax
  \kern\dimen@
  {#2}%
  \kern\dimen@
  \copy0 %                       second \perp
} 
\makeatother

\DeclareMathOperator{\E}{\textnormal{\mbox{E}}}

\usepackage[utf8]{inputenc}
\DeclareUnicodeCharacter{200E}{}

\usepackage{setspace}
\usepackage{cite}

\usepackage[stable]{footmisc}

\usepackage{rotating}

\makeatletter
\def\@hangfrom#1{\setbox\@tempboxa\hbox{{#1}}%
      \hangindent 0pt%\wd\@tempboxa
      \noindent\box\@tempboxa}
\makeatother

\makeatletter
\def\@seccntformat#1{\@ifundefined{#1@cntformat}%
   {\csname the#1\endcsname\quad}  % default
   {\csname #1@cntformat\endcsname}% enable individual control
}
\let\oldappendix\appendix %% save current definition of \appendix
\renewcommand\appendix{%
    \oldappendix
    \newcommand{\section@cntformat}{\appendixname~\thesection\quad}
}
\makeatother

\usepackage[absolute,showboxes]{textpos}

\TPMargin{5pt}

\newcommand{\copyrightstatement}{
    \begin{textblock}{0.84}(0.08,0.93)    % tweak here: {box width}(leftposition, rightposition)
         \noindent
         \footnotesize
         This draft manuscript presents work-in-progress. Comments are welcome at \href{mailto:idahabreh@hsph.harvard.edu}{idahabreh@hsph.harvard.edu}.
    \end{textblock}
}

\usepackage{thmtools, thm-restate}
\usepackage{datetime}
\def\paperversionmajor{9}
\def\paperversionminor{0}

\usepackage{sectsty}
\subsectionfont{\noindent\textbf\itshape}
\subsubsectionfont{\normalfont\itshape}

\usepackage{bibunits}

 \makeatletter
\def\hyper@natlinkstart#1{%
  \Hy@backout{#1}%
  \hyper@linkstart{cite}{cite.\@bibunitname.#1}%
  \def\hyper@nat@current{#1}%
}

\def\hyper@natlinkbreak#1#2{%
  \hyper@linkend#1\hyper@linkstart{cite}{cite.\@bibunitname.#2}%
}

\def\hyper@natanchorstart#1{%
  \hyper@anchorstart{cite.\@bibunitname.#1}%
}

\def\bibcite#1#2{%
  \@newl@bel{b}{#1}{\hyper@@link[cite]{}{cite.\@bibunitname.#1}{#2}}%
}%

\def\@lbibitem[#1]#2{%
  \@skiphyperreftrue
  \H@item[\hyper@anchorstart{cite.\@bibunitname.#2}%
  \@BIBLABEL{#1}\hyper@anchorend\hfill]%
  \@skiphyperreffalse
  \if@filesw
    \begingroup
      \let\protect\noexpand
      \immediate\write\@auxout{%
        \string\bibcite{#2}{#1}%
      }%
    \endgroup
  \fi
  \ignorespaces
}%

\def\@bibitem#1{%
  \@skiphyperreftrue\H@item\@skiphyperreffalse
  \hyper@anchorstart{cite.\@bibunitname.#1}\relax\hyper@anchorend
  \if@filesw
    \begingroup
      \let\protect\noexpand
      \immediate\write\@auxout{%
        \string\bibcite{#1}{\the\value{\@listctr}}%
      }%
    \endgroup
  \fi
  \ignorespaces
}%

\def\@citex[#1]#2{%
  \let\@citea\@empty
  \@cite{%
    \@for\@citeb:=#2\do{%
      \@citea
      \def\@citea{,\penalty\@m\ }%
      \edef\@citeb{\expandafter\@firstofone\@citeb}%
      \if@filesw
        \immediate\write\@auxout{\string\citation{\@citeb}}%
      \fi
      \@ifundefined{b@\@citeb}{%
        \mbox{\reset@font\bfseries ?}%
        \G@refundefinedtrue
        \@latex@warning{%
          Citation `\@citeb' on page \thepage \space undefined%
        }%
      }{%
        \hyper@natlinkstart{#2}%
            \hbox{\csname b@\@citeb\endcsname}%
        \hyper@natlinkend%
      }%
    }%
  }{#1}%
}%

\makeatother

\usepackage{subfiles}
\newcommand{\onlyinsubfile}[1]{#1}
\newcommand{\notinsubfile}[1]{}

\begin{document}

\renewcommand{\onlyinsubfile}[1]{}
\renewcommand{\notinsubfile}[1]{#1}

\begin{bibunit}[unsrt]

\copyrightstatement

\title{\textbf{Extending inferences from a cluster randomized trial to a target population} \vspace*{0.3in} }

%\copyrightstatement

\author[1-3]{Issa J. Dahabreh
%\thanks{Address for correspondence: Dr. Issa J. Dahabreh; Department of Epidemiology, Harvard T.H. Chan School of Public Health, Boston, MA 02115; email: \href{mailto:idahabreh@hsph.harvard.edu}{idahabreh@hsph.harvard.edu}; phone: +1 (617) 495‑1000.}
}
\author[1,2]{Sarah E. Robertson}
\author[4]{Jon A. Steingrimsson}
\author[5-7]{Stefan Gravenstein}
\author[8]{Nina Joyce}

\affil[1]{CAUSALab, Harvard T.H. Chan School of Public Health, Boston, MA}
\affil[2]{Department of Epidemiology, Harvard T.H. Chan School of Public Health, Boston, MA}
\affil[3]{Department of Biostatistics, Harvard T.H. Chan School of Public Health, Boston, MA}
\affil[4]{Department of Biostatistics, School of Public Health, Brown University, Providence, RI}
\affil[5]{Department of Health Services, Policy \& Practice, School of Public Health, Brown University, Providence, RI}
\affil[6]{Department of Medicine, Warren Alpert Medical School, Brown University, Providence, RI}
\affil[7]{Providence Veterans Administration Medical Center, Providence, RI}
\affil[8]{Department  Epidemiology, School of Public Health, Brown University, Providence, RI}

\maketitle{}
\thispagestyle{empty}

\newpage
\thispagestyle{empty}

\vspace*{1in}
{\Huge \centering Extending inferences from a cluster randomized trial to a target population \par }

\vspace{1in}
\noindent
\textbf{Running head:} Extending inferences from cluster randomized trials  

\vspace{0.3in}
\noindent
\textbf{Word count:} abstract= 92; main text=2770.

\vspace{0.3in}
\noindent
\textbf{Abbreviations that appear in the text:} LASSO = least absolute shrinkage and selection operator; LTC Focus = Long Term Care Focus; MDS = Minimum Data Set; OSCAR = Online Survey  Certification and Reporting; RHF = Residential History File.

\vspace{0.3in}
\noindent
\textbf{Abbreviations that appear in table legends:} AIPW = augmented inverse probability estimator, $\widehat \psi(a)$; IPW = inverse probability weighting estimator, $\widetilde \psi_{\text{\tiny w}}(a)$; LR (MLE) = logistic regression fit by maximum likelihood estimation methods; GAM = generalized additive model with logit link; LASSO = least absolute shrinkage and selection operator regression with logit link; EN = elastic net regularized regression with logit link; RF = random forest.

\clearpage
\thispagestyle{empty}

\vspace*{1in}

\begin{abstract}
\noindent
\linespread{1.7}\selectfont
We describe methods that extend (generalize or transport) causal inferences from cluster randomized trials to a target population of clusters, under a general nonparametric model that allows for arbitrary within-cluster dependence. We propose doubly robust estimators of potential outcome means in the target population that exploit individual-level data on covariates and outcomes to improve efficiency and are appropriate for use with machine learning methods. We illustrate the methods using a cluster randomized trial of influenza vaccination strategies conducted in 818 nursing homes nested in a cohort of 4,475 trial-eligible Medicare-certified nursing homes. \\

\vspace{0.2in}

\end{abstract}

% to do %%
% citations for cluster trials in intro

%%%%%%%%%%%%%%%%%%%%%%%%%%%%%%%%%%%%%%%%%%%%%%%%%%%%%%%%%%%%%%%%%%%%%%%%%%%%%%
\clearpage
\section{Introduction}
\setcounter{page}{1}
%%%%%%%%%%%%%%%%%%%%%%%%%%%%%%%%%%%%%%%%%%%%%%%%%%%%%%%%%%%%%%%%%%%%%%%%%%%%%%

When assessing interventions that are best applied to groups rather than individuals or when there is concern that exposure of one individual may affect the outcomes of other individuals in the same group, investigators can conduct a ``cluster randomized'' trial \cite{donner1994cluster}. In such a trial, selection for trial participation and treatment assignment is done at the cluster level, but covariate and outcome data are collected at the individual level. The randomized clusters are not necessarily representative of the target population of clusters where the treatment might be applied. Lack of representativeness is possible because the invitation to participate in the trial usually does not follow a formal sampling design and clusters invited to participate in the trial can decline the invitation. To the extent that randomized and non-randomized clusters differ in terms of covariates that are modifiers of the treatment effect, the average treatment effect in the trial may not apply to the entire population of trial-eligible clusters.

Methods for extending (generalizing or transporting \cite{hernan2016discussionkeiding, dahabreh2019commentaryonweiss}) inferences from an \emph{individually randomized} trial to a target population assume that individuals from the population are independently sampled and assigned treatment (e.g., \cite{dahabreh2018generalizing, buchanan2018generalizing, lesko2017practical}). These methods are not valid for cluster randomized trials where sampling and assignment occur at the cluster level and within-cluster dependence is expected \cite{balzer2019new, benitez2022comparative}. The few studies addressing generalizability for cluster randomized trials (e.g.,  \cite{omuircheartaigh2014, tipton2013improving, tipton2017implications}) have avoided these issues by aggregating all data at the cluster-level and using weighting or stratification methods appropriate for independent units. Such analyses, however, are problematic because aggregating information at the cluster level does not fully exploit individual-level information on covariates and outcomes. 

Here, we discuss analyses that extend inferences from cluster randomized trials to a target population, under a general nonparametric model that allows for arbitrary within-cluster dependence.  We propose doubly robust estimators that exploit individual-level data on covariates and outcomes and illustrate their use with data from a cluster randomized trial of influenza vaccination strategies in U.S. nursing homes.

%%%%%%%%%%%%%%%%%%%%%%%%%%%%%%%%%%%%%%%%%%%%%%%%%%%%%%%%%%%%%%%%%%%%%%%%%%%%%%
\section{Study design, data, and causal quantities of interest}\label{sec:data_causal_quant}
%%%%%%%%%%%%%%%%%%%%%%%%%%%%%%%%%%%%%%%%%%%%%%%%%%%%%%%%%%%%%%%%%%%%%%%%%%%%%%

\paragraph{Study design and data:} Consider the cluster version of the nested trial design for analyses extending inferences from a randomized trial to a target population \cite{dahabreh2019studydesigns}: among a cohort of trial-eligible clusters, a subset are invited and agree to participate in a randomized trial with cluster-level treatment assignment. In most cases, trial participation is not fully under investigator control. We index clusters in the cohort by $j \in \{1, \ldots, m\}$; the $j$th cluster has sample size $N_j$ and we allow the sample size to vary across clusters. Individuals in cluster $j$ are indexed by $i \in \{1, \ldots, N_j\}$. We use $S_j$ for the cluster-level indicator of participation in the trial; $S = 1$ for randomized clusters and $S = 0$ for non-randomized clusters. For all clusters -- both randomized and non-randomized -- we have data on cluster-level covariates, $X_j$, and a matrix of individual-level covariates for all individuals in the cluster, $\boldsymbol{W}_j$. If $p$ baseline covariates are collected from each individual in cluster $j$, then $\boldsymbol{W}_j$ has dimension $N_j \times p$. We use $A_j$ to denote the cluster level treatment assignment; we only consider finite sets of possible treatments, which we denote as $\mathcal A$. We assume that covariates $(X_j, \boldsymbol{W}_j)$ are measured at baseline, so that they cannot be affected by treatment. It is possible for participation in the trial and treatment assignment to depend on the covariate tuple $(X_j, \boldsymbol{W}_j)$: we expect that characteristics of the clusters or their members determine whether the clusters agree to participate in the trial and the treatment assignment probability in the trial may be conditional on covariates. Last, $\boldsymbol{Y}_j$ is the vector of individual-level outcomes, such that $\boldsymbol{Y}_j = (Y_{j,i} : i \in \{1, \ldots, N_j\})$ for each cluster $j$. We define the \emph{cluster-level average observed outcome} in cluster $j$, $\overbar{Y}_j$, as $\overbar{Y}_j = \frac{1}{N_j}\sum_{i = 1}^{N_j} Y_{j,i}.$

We assume independence across clusters, but we allow for arbitrary statistical dependence among individuals within each cluster. Such dependence can occur because of (1) \emph{shared exposures:} individuals share measured and unmeasured cluster-level factors; (2) \emph{contagion:} outcomes may be contagious, so that $Y_{j,i}$ may depend on $Y_{j,i^\prime}$ for any two individuals, $i$ and $i^\prime$, in cluster $j$; (3) \emph{covariate interference:} one individual's covariates may affect another individual's outcomes, so that $Y_{j,i}$ may depend on $X_{j,i^\prime}$ for any two individuals, $i$ and $i^\prime$, in cluster $j$; or (4) \emph{treatment-outcome interference:} one individual's treatment assignment may affect another individuals outcome, so that $Y_{j,i}$ may depend on $A_{j,i^\prime}$ for any two individuals, $i$ and $i^\prime$, in cluster $j$.

We collect data on baseline covariates from the cohort of trial-eligible clusters and view the cohort as a simple random sample from that target population of clusters. Treatment and outcome data need only be collected from clusters participating in the trial. More formally, the observed data may be independent and identically distributed realizations of the random tuple $O_j = (X_j, \boldsymbol{W}_j, S_j, A_j, \boldsymbol{Y}_j)$, $j \in \{1, \ldots, m\}$, but our results also apply to data $(X_j, \boldsymbol{W}_j, S_j, S_j \times A_j, S_j \times \boldsymbol{Y}_j)$, $j \in \{1, \ldots, m\}$ when treatment and outcome information is unavailable from non-randomized clusters.

\paragraph{Causal quantities:} Let $Y_{j,i}^a$ be the potential outcome for individual $i$ in cluster $j$ under intervention to assign treatment $a \in \mathcal A$ \cite{rubin1974, robins2000d} and let $\boldsymbol{Y}^a_j$ denote the vector of potential (counterfactual) outcomes in cluster $j$ under intervention to assign treatment $a \in \mathcal A$, such that $\boldsymbol{Y}^a_j = (Y_{j,i}^a : i \in \{1, \ldots, N_j\})$. We define the \emph{cluster-level average counterfactual outcome} in cluster $j$, $\overbar{Y}^{a}_j$, as $\overbar{Y}^{a}_j = \frac{1}{N_j}\sum_{i = 1}^{N_j} Y_{j,i}^a$ \cite{balzer2019new, benitez2022comparative}.

The main causal quantity of interest is the potential outcome mean in the target population, $\E\left[\overbar{Y}^{a}\right] = \E\left[\frac{1}{N_j}\sum_{i = 1}^{N_j} Y_{j,i}^a \right].$ This potential outcome mean will be different from the potential outcome mean in the population of randomized clusters, that is $\E\left[\overbar{Y}^{a}\right] \neq \E\left[\overbar{Y}^{a} | S = 1\right]$, when factors which affect the outcome are differentially distributed between clusters that participate in the trial and those that do not. Treatment effects in the target population can be defined using these potential outcome means. For example, we define the average treatment effect in the target population comparing treatments $a \in \mathcal A$ and $a^\prime \in \mathcal A$, as $\E\left[\overbar{Y}^{a} - \overbar{Y}^{a^\prime} \right] = \E\left[\overbar{Y}^{a}\right] - \E\left[ \overbar{Y}^{a^\prime} \right]$.

Another causal quantity of potential interest is the average treatment effect among the non-randomized subset of the target population, $\E\left[\overbar{Y}^{a} | S = 0 \right]$. In the main text of this paper we focus our attention on $\E\left[\overbar{Y}^{a}  \right]$; we collect all results about $\E\left[\overbar{Y}^{a} | S = 0 \right]$ in the Appendix.

%%%%%%%%%%%%%%%%%%%%%%%%%%%%%%%%%%%%%%%%%%%%%%%%%%%%%%%%%%%%%%%%%%%%%%%%%%%%%%
\section{Identification}
%%%%%%%%%%%%%%%%%%%%%%%%%%%%%%%%%%%%%%%%%%%%%%%%%%%%%%%%%%%%%%%%%%%%%%%%%%%%%%

\paragraph{Identifiability conditions:} The following conditions are sufficient to identify the potential outcome means $\E[\overbar{Y}^{a}]$: \emph{A1. Consistency of cluster-level average counterfactual outcomes:} if $A_j = a$, then $\boldsymbol{Y}^{a}_j = \boldsymbol{Y}_j$ for every $j \in \{1, \ldots, m\}$ and every $a \in \mathcal A$. \emph{A2. Conditional exchangeability over $A$ in the cluster randomized trial:} $\boldsymbol{Y}^{a} \indep A | X, \boldsymbol{W}, S = 1$. \emph{A3. Positivity of treatment assignment probability in the trial:} $\Pr[A = a | X = x, \boldsymbol{W} = \boldsymbol{w}, S = 1] > 0$ for every $a$, and every $x$ and $\boldsymbol{w}$ with positive density in the trial. \emph{A4. Conditional exchangeability over $S$:} $\boldsymbol{Y}^{a} \indep S | X, \boldsymbol{W}$. \emph{A5. Positivity of trial participation:} $\Pr[S = 1 | X = x, \boldsymbol{W} = \boldsymbol{w}] > 0$ for every $x$ and $\boldsymbol{w}$ with positive density in the target population.

\paragraph{Identification:} In the Appendix, we show that under the above conditions, $\E\left[\overbar{Y}^{a}\right]$ is identified by $\psi(a) \equiv \E \Big[ \E\left[\overbar{Y} | X, \boldsymbol{W}, S = 1, A = a \right] \Big]$, where again the outer expectation is over the distribution of the target population. The average treatment effect comparing interventions $a$ and $a^\prime$ is identified because $\E\left[\overbar{Y}^{a} - \overbar{Y}^{a^\prime}\right] = \psi(a) - \psi(a^\prime)$.

Note that the exchangeability conditions and this identification result can be obtained using graphical causal models (e.g., directed acyclig graphs with selection nodes \cite{pearl2014, bareinboim2016causalfusion} or single world intervention graphs \cite{richardson2013single} where trial participation is viewed as an intervention \cite{dahabreh2019identification}). Here, we take the exchangeability conditions as our starting point (i.e., as primitive conditions), to focus on aspects related to data analysis. In the Appendix, we also show that the average treatment effect (but not its component potential outcome means) is identifiable under weaker assumptions of exchangeability in measure \cite{vanderWeele2012,dahabreh2018generalizing}.

%%%%%%%%%%%%%%%%%%%%%%%%%%%%%%%%%%%%%%%%%%%%%%%%%%%%%%%%%%%%%%%%%%%%%%%%%%%%%%
\section{Estimation and inference} \label{sec:estimation}
%%%%%%%%%%%%%%%%%%%%%%%%%%%%%%%%%%%%%%%%%%%%%%%%%%%%%%%%%%%%%%%%%%%%%%%%%%%%%%

\paragraph{Estimation:} The structure of $\psi(a)$ combined with prior work on efficient estimation with individually randomized trials \cite{dahabreh2018generalizing} suggest the following augmented inverse probability of participation weighting estimator
\begin{equation*}\label{eq:estimator_dr}
    \begin{split}
    \widehat{\psi}(a) &= \dfrac{1}{m} \sum\limits_{j = 1}^{m} \Bigg\{ \dfrac{I(S_j = 1, A_j = a)}{\widehat p(X_j, \boldsymbol{W}_j) \widehat e_a(X_j, \boldsymbol{W}_j)}  \Big\{ \overbar{Y}_j - \widehat g_a(X_j, \boldsymbol{W}_j) \Big\} + \widehat g_a(X_j, \boldsymbol{W}_j) \Bigg\},
    \end{split}
\end{equation*}
where $\widehat p(X, \boldsymbol{W})$ is an estimator for $\Pr[S = 1 | X, \boldsymbol{W}] $; $\widehat e_a(X, \boldsymbol{W})$ is an estimator for $\Pr[ A  = a | X, \boldsymbol{W}, S = 1]$ (alternatively, the true randomization probability can be used); and $\widehat g_a(X, \boldsymbol{W})$ is an estimator for $\E\left[ \overbar{Y} | X, \boldsymbol{W}, S = 1, A = a \right]$.  

The estimators $\widehat p(X, \boldsymbol{W})$ and, if desired, $\widehat e_a(X, \boldsymbol{W})$, can be easily obtained using cluster-level data. In contrast, estimating $\E\left[ \overbar{Y} | X, \boldsymbol{W}, S = 1, A = a \right]$ using only cluster-level data would require averaging the outcomes in each cluster to obtain $\overbar{Y}$, failing to exploit the availability of individual-level information on individual-level outcomes, as well as covariates in the cluster randomized trial. Instead, we modify the strategy in \cite{balzer2019new} for use in the context of analyses extending causal inferences to a target population: We begin by specifying and fitting a working regression model for the conditional expectation of the individual-level outcome $Y_{j,i}$ given cluster level covariates $X_j$ and the individual's covariates $W_{j,i}$ (not the entire matrix $\boldsymbol{W}_{j}$); the regression can be fit separately by trial arm, to allow for effect modification. Next, we obtain estimated values from the fitted model on all individuals in the data, regardless of trial participation status. We denote these predictions as $\widehat h_a(X_{j}, W_{j,i})$. Last, we obtain estimates $\widehat g_a(X_j, \boldsymbol{W}_j)$ by averaging the predictions over individuals in each cluster, $ \widehat g_a(X_j, \boldsymbol{W}_j) = \frac{1}{N_j} \sum_{i = 1}^{N_j} \widehat h_a(X_{j,i}, W_{j,i})$. 

In the Appendix, we derive an asymptotic representation for $\widehat \psi(a)$ when $\Pr[S = 1 | X, \boldsymbol{W}] $ and $\E\left[ \overbar{Y} | X, \boldsymbol{W}, S = 1, A = a \right]$ need to be estimated using (possibly misspecified) statistical models. Here, we note that the estimator $\widehat \psi(a)$ is doubly robust, in the sense that it is consistent and asymptotically normal when either $\widehat p(X, \boldsymbol{W})$ or $\widehat g_a(X, \boldsymbol{W})$ are consistent estimators, but not necessarily both. This robustness property is practically important because, in general, dependence among units makes correct specification of models for estimating $\widehat g_a(X, \boldsymbol{W})$ challenging. We view these ``working'' models as estimation tools aimed primarily at increasing efficiency by using individual-level outcomes on covariates and outcomes. Of course, should more detailed information be available (e.g., about networks of contacts within each cluster), we could use it to better estimate  $\widehat g_a(X, \boldsymbol{W})$. 

In the Appendix, we describe a g-formula estimator and a non-augmented inverse probability weighting estimator, both of which can be viewed as special (non-robust) versions of $\widehat \psi(a)$.

\paragraph{Inference:} Following \cite{balzer2019new}, we estimate the variance of the sampling distribution of $\widehat \psi(a)$ as 
\begin{equation*}
\widehat \sigma^2_{\widehat \psi(a)} = \dfrac{1}{m} \widehat{\mbox{Var}} \Big[ \widehat{\mathit{\Psi}}^1_j(a) \Big],
\end{equation*}
where $ \widehat{\mbox{Var}} \Big[ \widehat{\mathit{\Psi}}^1_j(a) \Big] $ is the estimated variance of the influence curve, $$ \widehat{\mathit{\Psi}}^1_j(a) = \dfrac{I(S_j = 1, A_j = a)}{\widehat p(X_j, \boldsymbol{W}_j) \widehat e_a(X_j, \boldsymbol{W}_j)} \Big\{ \overbar{Y}_j - \widehat g_a(X_j, \boldsymbol{W}_j) \Big\} + \widehat g_a(X_j, \boldsymbol{W}_j) - \widehat \psi(a).$$ The sampling variance can be used to obtain a $(1 - \alpha)\%$ confidence interval as $\Big( \widehat \psi(a) \pm z_{1-\alpha/2} \times \widehat \sigma_{\widehat \psi(a)} \Big),$ where $z_{1-\alpha/2}$ is the $(1-\alpha/2)$ quantile of the standard normal distribution. This confidence interval is easy to compute and simulation studies in other contexts \cite{balzer2019new} suggest that it performs well. Bootstrap methods can also be used, as an alternative \cite{efron1994introduction}.

%%%%%%%%%%%%%%%%%%%%%%%%%%%%%%%%%%%%%%%%%%%%%%%%%%%%%%%%%%%%%%%%%%%%%%%%%%%%%%
\section{Application of the methods} \label{sec:application}
%%%%%%%%%%%%%%%%%%%%%%%%%%%%%%%%%%%%%%%%%%%%%%%%%%%%%%%%%%%%%%%%%%%%%%%%%%%%%%

\paragraph{Study design and data:} We obtained data from a 2x2 factorial multi-facility cluster randomized trial conducted over the 2013-2014 influenza season (October 2013 through May 2014) that compared high-dose versus standard-dose influenza vaccines for eligible residents and standard-dose vaccination versus standard-of-care for nursing home staff \cite{gravenstein2016cluster, gravenstein2017comparative}. 

To be eligible for the trial, nursing homes had to be free-standing facilities (i.e., not hospital based) within 50-miles of a CDC-reporting city, have a minimum of 50 long-stay residents, and at least 80\% of residents aged 65 years or older. Within each facility, residents were candidates for the compared vaccines if they were 65 or older, long-stay residents and had no contraindications for high-dose influenza vaccine. Consistent with the analytic approach in the trial analyses, we restricted our analyses to Medicare beneficiaries only, because the exact date of death is not reliably captured for non-Medicare-eligible nursing home residents. Because treatment was assigned at the nursing-home level, we used data from all residents (not only the candidates for the compared vaccines).

We obtained data on all Medicare-certified nursing homes operating at the time of trial initiation from the Long Term Care (LTC) Focus database, which combines information from the Online Survey Certification and Reporting (OSCAR) survey, the nursing home Minimum Data Set (MDS) 3.0, and the Medicare Residential History File (RHF). We obtained individual-level covariates from the MDS 3.0 and date of death information from the Medicare Vital Status File. All individual-level covariates were obtained from the most recent assessment prior to the start of the trial.

The variables used to define trial eligibility were all derived from the LTC Focus database. Thus, using the same data, we were able to identify all Medicare-certified trial-eligible nursing homes in the U.S. at the time of trial initiation (i.e., all homes that could potentially have participated in the trial). We linked the trial dataset with the dataset of all trial-eligible nursing homes, nesting the cluster randomized trial in the cohort of trial-eligible nursing homes \cite{dahabreh2019studydesigns}. Thus, we had access to nursing home- and resident-level covariate data from all trial-eligible nursing homes, along with nursing home-level treatment assignment data and resident-level outcome data for all Medicare beneficiaries residing in nursing homes that actually participated in the trial. 

We used death from any cause as the outcome of interest. The followup time was brief (1 respiratory season) and there was no censoring for the outcome of death (the Medicare Vital Status File captures death data no matter where the death occurred, e.g., in a nursing home or in the community). Thus, cumulative incidence proportions and their differences are reasonable measures of incidence and effect, respectively.

\paragraph{Analysis methods:} We used the methods described earlier in the paper to estimate potential outcome means for the four vaccination strategies in the target population of trial-eligible nursing homes.

When modeling trial participation and treatment assignment we used cluster-level covariates as predictors (whether the facility was part of a chain, whether it had for-profit status, the total number of beds, the registered nurse (RN)/nurse ratio, the average acuity index measure of the level of care required  by nursing home residents, and the \% of residents with pneumonia, congestive heart failure, hospitalization in the prior year, and vaccinated in the prior year). When modeling the outcome, we used death as the response variable (at the individual-level) and predictors at the facility-level (same as for the participation model) and at the individual-level (age, Alzheimer's disease, diabetes, pneumonia, congestive heart failure, hypertension, cirrhosis, chronic kidney disease, presence of an indwelling catheter, use of ventilator, atrial fibrillation, coronary artery disease, receipt of chemotherapy, receipt of radiation, dialysis). 

We estimated the conditional probability of trial participation (nursing home-level) and  death (individual-level, separately for each treatment group) using different methods to illustrate that our approach can easily accommodate them. One set of analyses used logistic regression models estimated by maximum likelihood methods. For comparison, we also estimated probabilities using (1) generalized additive models \cite{hastie1990monographs}; (2) least absolute shrinkage and selection operator (LASSO) regression \cite{tibshirani1996regression}; (3) elastic net regularized regression \cite{zou2005regularization}; and (4) random forests \cite{breiman2001random}.

Regardless of how we estimated the conditional probability of trial participation and death, we always estimated the probability of treatment among trial participants \cite{dahabreh2019relation} using logistic regression estimated by maximum likelihood methods (the trial was marginally randomized, so the parametric logistic regression model is correctly specified and there is no need to use machine learning methods to estimate the probability of participation). 

Fitted values from each approach were combined using $\widehat \psi(a)$. For comparison, we also obtained estimates from a non-augmented inverse probability weighting estimator applied to cluster-level data (see Appendix for details; we obtained estimates from this estimator only using parametric logistic regression models \cite{chernozhukov2018double}). Last, we also obtained a trial-only estimator using ordinary least squares regression to fit a saturated linear probability model on individual-level data.

For augmented inverse probability weighting estimators, we obtained 95\% influence curve-based intervals as described earlier in this paper. For the non-augmented inverse probability weighting estimator, we obtained confidence intervals using the robust variance estimator with cluster-level data. For the trial-only estimator, we obtained confidence intervals using cluster- and heteroskedasticity-robust standard errors \cite{cameron2015practitioner}.

\paragraph{Results:} We had data from 818 randomized nursing homes with a total of 82,604 residents. The randomized trial was nested within a cohort of 4,475 trial-eligible nursing homes with a total of  523,764 residents; trial-participating homes were 18.3\% of trial-eligible homes (and included 15.8\% of all residents in trial-eligible homes). 

Table \ref{table:estimation_results} presents results from analyses using different estimators and different methods to estimate working models. Overall, the augmented inverse probability weighting estimator produced similar point estimates for the potential outcome mean under each treatment across all approaches for estimating the working models. These point estimates were fairly similar to those of the non-augmented inverse probability weighting estimator for all treatments except the control strategy of standard dose vaccination for residents and standard-of-care vaccination for staff. For that strategy, the non-augmented estimator produced estimates lower by about 2\%  and the point estimate from the non-augmented estimator was lower than the lower limit of the 95\% confidence interval of all augmented estimators. Confidence intervals were narrowest when using the augmented inverse probability weighting estimator with the random forest algorithm. More generally, confidence intervals using the augmented estimator were of equal or narrower width compared to the non-augmented estimator; the difference was most pronounced for the control treatment. Point estimates for generalizability analyses were largely similar to those from the randomized trial, suggesting that the trial-only estimates were applicable to the broader population of trial-eligible nursing homes for the outcome of all-cause death.

%%%%%%%%%%%%%%%%%%%%%%%%%%%%%%%%%%%%%%%%%%%%%%%%%%%%%%%%%%%%%%%%%%%%%%%%%%%%%%
\section{Discussion}
%%%%%%%%%%%%%%%%%%%%%%%%%%%%%%%%%%%%%%%%%%%%%%%%%%%%%%%%%%%%%%%%%%%%%%%%%%%%%%

We described methods for extending causal inferences from cluster randomized trials to a well-defined target population of clusters. Our main contribution is to show how recent work on generalizability and transportability methods can be extended to the cluster trial setting under a general causal model that allows for arbitrary within-cluster dependence, by leveraging novel methods for the analysis of clustered data \cite{balzer2019new, benitez2022comparative}. Allowing for within-cluster dependence is particularly important when studying ``dependent happenings'' \cite{halloran1995causal, halloran2016dependent} as is typically the case in vaccine research (e.g., due to ``herd immunity,'' or in the context of disease outbreaks and the implementation of measures for their control).

The methods we propose have several advantages compared with previous work on generalizability with cluster randomized trials (e.g., \cite{omuircheartaigh2014, tipton2013improving, tipton2017implications}): first, our estimators offer some degree of robustness to model misspecification \cite{bang2005}; second, the double robustness property allows us to exploit individual-level data on covariates and outcomes leading to improvements in efficiency compared to approaches that only use cluster-level data \cite{balzer2019new}; third, our estimators can be easily combined with flexible data-adaptive approaches for estimating working models (e.g., machine learning methods) \cite{chernozhukov2018double}. 

Future work can extend these methods to address use with incomplete outcome data (e.g., due to drop-out and other causes of loss-to-followup), competing risks (e.g., for outcomes other than mortality or to investigate cause-specific mortality), or different causal quantities of interest (e.g., counterfactual survival curves). Future work can also extend the methods to take advantage of additional structural information about within-cluster dependence (e.g., contact networks).

\iffalse 

\section{Acknowledgement}

This section has been temporarily removed from the manuscript (for peer review).

SG reports grants, personal fees, and nonfinancial support from Genentech, Sanofi and Seqirus; personal fees from Longevoron, Janssen, the American Geriatrics Society, and the Gerontological Society of America; and grants and/or contracts from the National Institutes of Health (NIH), Pfizer, Seqirus, Sanofi and Janssen.

This work was supported in part by Patient-Centered Outcomes Research Institute (PCORI) award ME-1502-27794 (Dahabreh) and Agency for Healthcare Research and Quality (AHRQ) National Research Service Award T32AGHS00001 (Robertson).  \\
The data analyses in this paper used data obtained from a completed clinical trial (registered with ClinicalTrials.gov, number NCT01815268) funded by an investigator-initiated grant from Sanofi Pasteur (GRC75-EXT). Insight Therapeutics served as the clinical research organization for study management and data collection. The authors thank H. Edward Davison, PharmD, MPH and Lisa Han, MPH (both at Insight Therapeutics) for their contribution to the original clinical trial.\\
The content of this paper is solely the responsibility of the authors and does not necessarily represent the official views of PCORI, its Board of Governors, the PCORI Methodology Committee, AHRQ, Sanofi Pasteur, or Insight Therapeutics.

\fi

%%%%%%%%%%%%%%%%%%%%%%%%%%%%%%%%%%%%%%%%%%%%%%%%%%%%%%%%%%%%%%%%%%%%%%%%%%%%%%
% BIBLIOGRAPHY
%%%%%%%%%%%%%%%%%%%%%%%%%%%%%%%%%%%%%%%%%%%%%%%%%%%%%%%%%%%%%%%%%%%%%%%%%%%%%%
\clearpage
\bibliographystyle{unsrt}
%\bibliography{bibliography}
\putbib[bibliography]
%%%%%%%%%%%%%%%%%%%%%%%%%%%%%%%%%%%%%%%%%%%%%%%%%%%%%%%%%%%%%%%%%%%%%%%%%%%%%%

%%%%%%%%%%%%%%%%%%%%%%%%%%%%%%%%%%%%%%%%%%%%%%%%%%%%%%%%%%%%%%%%%%%%%%%%%%%%%%
% VERSIONING 
%%%%%%%%%%%%%%%%%%%%%%%%%%%%%%%%%%%%%%%%%%%%%%%%%%%%%%%%%%%%%%%%%%%%%%%%%%%%%%

\ddmmyyyydate %redefine \today format
\newtimeformat{24h60m60s}{\twodigit{\THEHOUR}.\twodigit{\THEMINUTE}.32}
\settimeformat{24h60m60s}
\begin{center}
\vspace{\fill}\ \newline
\textcolor{black}{{\tiny $ $generalizability\_clusters\_estimation, $ $ }
{\tiny $ $Date: \today~~ \currenttime $ $ }
{\tiny $ $Revision: \paperversionmajor.\paperversionminor $ $ }}
\end{center}
%%
%%

%%%%%%%%%%%%%%%%%%%%%%%%%%%%%%%%%%%%%%%%%%%%%%%%%%%%%%%%%%%%%%%%%%%%%%%%%%%%%%
\clearpage
\section*{Table}
%%%%%%%%%%%%%%%%%%%%%%%%%%%%%%%%%%%%%%%%%%%%%%%%%%%%%%%%%%%%%%%%%%%%%%%%%%%%%%

\begin{table}[h!]

\caption{Results for different estimators and different algorithms for estimating the working models in a cluster randomized trial of influenza vaccination strategies.}\label{table:estimation_results}
\centering
{
\renewcommand{\arraystretch}{1.30}
\resizebox{\textwidth}{!}{
\begin{tabular}{|l|l|l|l|l|l|l|}

\hline
\multirow{2}{*}{\textbf{Estimator}} & \multirow{2}{*}{\textbf{Working model}} & \multicolumn{4}{c|}{\textbf{Vaccination strategy}}                                                          \\ \cline{3-6} 
                                    &                                                                                               & \multicolumn{1}{c|}{$a=1$} & \multicolumn{1}{c|}{$a=2$} & \multicolumn{1}{c|}{$a=3$} & \multicolumn{1}{c|}{$a=4$} \\ \hline
Trial-only                          & Unadjusted                                                                         & 20.82 (20.08, 21.57)       &     20.73   (19.99, 21.48)  & 20.76 (20.06, 21.46)                               &  21.15 (20.30, 22.01)    \\ \hline
IPW                        & LR (MLE)                                                                                      &   20.57  (19.56, 21.57)   &  20.74 (19.28, 22.21) &           21.18 (20.18, 22.18) &   19.77 (17.49, 22.05)  \\ \hline
AIPW                                & LR (MLE)                                                                                      &   21.36   (20.35, 22.36)                               &       21.27   (20.08, 22.47)                           &    21.50   (20.58, 22.42)                              &     21.81   (20.52, 23.11)                                        \\ \hline
AIPW                                & GAM                                                                                         &   21.35   (20.49, 22.21)                               &          20.77   (19.74, 21.80)  &    21.42   (20.57, 22.28)   &     21.74   (20.47, 23.02)                                       \\ \hline
AIPW                                & LASSO                                                                                   &  21.26   (20.17, 22.34)  &   21.01   (19.81, 22.20)   &             21.55   (20.66, 22.45)   &   21.72   (20.63, 22.82)    \\ \hline
AIPW                                & EN                                                                                 &    21.38   (20.39, 22.38)   &   21.10   (20.01, 22.19)   &       21.57   (20.69, 22.45) &    21.74   (20.68, 22.79)   \\ \hline
AIPW                                & RF                                                                                           &  21.86   (21.41, 22.31)   &    21.73   (21.27, 22.18)  &    21.75   (21.32, 22.19)    &    21.98   (21.54, 22.42)    \\ \hline
\end{tabular}
}
}
\caption*{Results are reported as point estimates of the cumulative incidence proportion (95\% confidence intervals). \\
Treatment strategies are denoted as follows: $a=1$, high-dose resident vaccination and standard-dose staff vaccination; $a=2$, high-dose resident vaccination and standard-of-care staff vaccination; $a=3$, standard-dose resident vaccination and standard-dose staff vaccination; and $a=4$, standard-dose resident vaccination and standard-of-care staff vaccination. 
\\
AIPW = augmented inverse probability estimator, $\widehat \psi(a)$; IPW = inverse probability weighting estimator, $\widetilde \psi_{\text{\tiny w}}(a)$; LR (MLE) = logistic regression fit by maximum likelihood estimation methods; GAM = generalized additive model with logit link; LASSO = least absolute shrinkage and selection operator regression with logit link; EN = elastic net regularized regression with logit link; RF = random forest. The formula for  $\widetilde \psi_{\text{\tiny w}}(a)$ is given in the Appendix.}
\end{table}

\end{bibunit}

\maketitle{}

\begin{bibunit}[unsrt]

\onlyinsubfile{

\title{\emph{Appendix to} \\ \textbf{Extending inferences from a cluster randomized trial to a target population} \vspace*{0.3in} }

\copyrightstatement

\author[1-3]{Issa J. Dahabreh
%thanks{Address for correspondence: Dr. Issa J. Dahabreh; Department of Epidemiology, Harvard T.H. Chan School of Public Health, Boston, MA 02115; email: \href{mailto:idahabreh@hsph.harvard.edu}{idahabreh@hsph.harvard.edu}; phone: +1 (617) 495‑1000.}
}
\author[1,2]{Sarah E. Robertson}
\author[4]{Jon A. Steingrimsson}
\author[5-7]{Stefan Gravenstein}
\author[8]{Nina Joyce}

\affil[1]{CAUSALab, Harvard T.H. Chan School of Public Health, Boston, MA}
\affil[2]{Department of Epidemiology, Harvard T.H. Chan School of Public Health, Boston, MA}
\affil[3]{Department of Biostatistics, Harvard T.H. Chan School of Public Health, Boston, MA}
\affil[4]{Department of Biostatistics, School of Public Health, Brown University, Providence, RI}
\affil[5]{Department of Health Services, Policy \& Practice, School of Public Health, Brown University, Providence, RI}
\affil[6]{Department of Medicine, Warren Alpert Medical School, Brown University, Providence, RI}
\affil[7]{Providence Veterans Administration Medical Center, Providence, RI}
\affil[8]{Department  Epidemiology, School of Public Health, Brown University, Providence, RI}

\maketitle{}
\thispagestyle{empty}

}

\maketitle{}

%%%%%%%%%%%%%%%%%%%%%%%%%%%%%%%%%%%%%%%%%%%%%%%%%%%%%%%%%%%%
%%%%%%%%%%%%%%%%%%%%%%%%%%%%%%%%%%%%%%%%%%%%%%%%%%%%%%%%%%%%
%%%%%%%%%%%%%%%%%%%%%%%%%%%%%%%%%%%%%%%%%%%%%%%%%%%%%%%%%%%%
%%%%%%%%%%%%%%%%%%%%%%%%%%%%%%%%%%%%%%%%%%%%%%%%%%%%%%%%%%%%
\clearpage
\setcounter{page}{1}
\appendix
%%%%%%%%%%%%%%%%%%%%%%%%%%%%%%%%%%%%%%%%%%%%%%%%%%%%%%%%%%%%
%%%%%%%%%%%%%%%%%%%%%%%%%%%%%%%%%%%%%%%%%%%%%%%%%%%%%%%%%%%%
%%%%%%%%%%%%%%%%%%%%%%%%%%%%%%%%%%%%%%%%%%%%%%%%%%%%%%%%%%%%
%%%%%%%%%%%%%%%%%%%%%%%%%%%%%%%%%%%%%%%%%%%%%%%%%%%%%%%%%%%%

%%%%%%%%%%%%%%%%%%%%%%%%%%%%%%%%%%%%%%%%%%%%%%%%%%%%%%%%%%%%
%%%%%%%%%%%%%%%%%%%%%%%%%%%%%%%%%%%%%%%%%%%%%%%%%%%%%%%%%%%%
\section{Identification and estimation of \texorpdfstring{$\E\left[\overbar{Y}^{a}\right]$}{e}}\label{appendix_A}
\renewcommand{\theequation}{A.\arabic{equation}}
\setcounter{equation}{0}
%%%%%%%%%%%%%%%%%%%%%%%%%%%%%%%%%%%%%%%%%%%%%%%%%%%%%%%%%%%%
%%%%%%%%%%%%%%%%%%%%%%%%%%%%%%%%%%%%%%%%%%%%%%%%%%%%%%%%%%%%

Throughout this Appendix, the observed data are $O = (X, \boldsymbol{W}, S , A, \boldsymbol{Y})$, drawn from some unspecified population distribution of clusters, $P$, that is, $O \sim P$. Unless otherwise stated, we assume that $P$ belongs in a completely unspecified class of distributions $\mathcal M$, $P \in \mathcal M$. The density of $P$ factorizes as $$p_{O}(o) = p_{\boldsymbol{Y}|X, \boldsymbol{W},S,A}(\boldsymbol{y}|x,\boldsymbol{w},s,a) p_{A|X,\boldsymbol{W},S}(a|x,\boldsymbol{w},s) p_{X,\boldsymbol{W}|S}(x,\boldsymbol{w}|s) p_{S}(s).$$

%%%%%%%%%%%%%%%%%%%%%%%%%%%%%%%%%%%%%%%%%%%%%%%%%%%%%%%%%%%%
\subsection{Identification}
%%%%%%%%%%%%%%%%%%%%%%%%%%%%%%%%%%%%%%%%%%%%%%%%%%%%%%%%%%%%

\begin{proposition}
Under identifiability conditions A1 through A5 in the main text,  the expectation of the average potential outcome in the target population, $ \E\left[\overbar{Y}^{a}\right] $, is identified by $$\psi(a) \equiv \E \Big[ \E\left[\overbar{Y} | X, \boldsymbol{W}, S = 1, A = a \right] \Big].$$
\end{proposition}

Starting with the causal quantity of interest,
\begin{equation*}
    \begin{split}
        \E\left[\overbar{Y}^{a}\right] &= \E \Big[ \E\left[\overbar{Y}^{a} | X, \boldsymbol{W} \right] \Big] \\
            &= \E \Big[ \E\left[\overbar{Y}^{a} | X, \boldsymbol{W}, S = 1 \right] \Big] \\
            &= \E \Big[ \E\left[\overbar{Y}^{a} | X, \boldsymbol{W}, S = 1, A = a \right] \Big] \\
            &= \E \Big[ \E\left[\overbar{Y} | X, \boldsymbol{W}, S = 1, A = a \right] \Big] \\
            &= \psi(a),
    \end{split}
\end{equation*}
where the first step follows from the law of total expectation, the second by condition \emph{A4}, the third by condition \emph{A2}, the fourth by condition \emph{A1}, the last by the definition of $\psi(a)$; quantities are well-defined because of the positivity conditions \emph{A3} and \emph{A5}. We note that identification of $\E[\overbar{Y}^a ]$ is also possible under weaker conditions of exchangeability in mean rather than in distribution (e.g., see \cite{dahabreh2018generalizing} for arguments in the individually-randomized trial case that can be modified for the cluster randomized trial case).

\paragraph{Identification of the average treatment effect under weaker conditions:} As in earlier work \cite{dahabreh2018generalizing}, if the causal quantity of interest is only the average treatment effect in the target population (but not the treatment-specific potential outcome means), then identification is possible under weaker conditions. Specifically, consider the following condition, as an alternative to condition \emph{A4} from the main text:

\vspace{0.1in}
\noindent
\emph{A4$^*$. Conditional exchangeability in measure over $S$:} for treatments $a \in \mathcal A$ and $a^\prime \in \mathcal A$, $\E[\overbar{Y}^a - \overbar{Y}^{a^\prime} | X = x, \boldsymbol{W} = \boldsymbol{w}] = \E[\overbar{Y}^a - \overbar{Y}^{a^\prime} | X = x, \boldsymbol{W} = \boldsymbol{w}, S = 1]$ for all $x$ and $\boldsymbol{w}$ with positive density in the target population.

\begin{proposition}
Under conditions A1 through A3, A4$^*$, and A5 the average treatment effect in the target population comparing treatments $a$ and $a^\prime$ is identifiable by $$\psi(a, a^\prime) \equiv \E \Big[ \E\Big[\overbar{Y} | X, \boldsymbol{W}, S = 1, A = a \Big] - \E\Big[\overbar{Y} | X, \boldsymbol{W}, S = 1, A = a^\prime \Big] \Big].$$
\end{proposition}

Starting with the causal quantity of interest,
\begin{equation*}
    \begin{split}
        \E\Big[\overbar{Y}^{a} - \overbar{Y}^{a^\prime} \Big] &= \E \Big[ \E\left[\overbar{Y}^{a} -  \overbar{Y}^{a^\prime}| X, \boldsymbol{W} \right] \Big] \\
            &= \E \Big[ \E\left[\overbar{Y}^{a} - \overbar{Y}^{a^\prime} | X, \boldsymbol{W}, S = 1 \right] \Big] \\
            &= \E \Big[ \E\Big[\overbar{Y}^{a} | X, \boldsymbol{W}, S = 1, A = a \Big] - \E\Big[\overbar{Y}^{a^\prime} | X, \boldsymbol{W}, S = 1, A = a^\prime \Big] \Big] \\
            &= \E \Big[ \E\Big[\overbar{Y} | X, \boldsymbol{W}, S = 1, A = a \Big] - \E\Big[\overbar{Y} | X, \boldsymbol{W}, S = 1, A = a^\prime \Big] \Big] \\
            & \psi(a, a^\prime),
    \end{split}
\end{equation*}
where the first step follows from the law of total expectation, the second by condition \emph{A4}$^*$, the third by condition \emph{A2}, the fourth by condition \emph{A1}, and the last by the definition of $\psi(a, a^\prime)$; quantities are well-defined because of the positivity conditions \emph{A3} and \emph{A5}.

\paragraph{Identification without treatment and outcome data among non-randomized clusters:} A careful inspection of the results in this section of the Appendix reveals that treatment and outcome information from non-randomized clusters is not necessary for identification; that is to say, data on $(X, \boldsymbol{W}, S, S \times A, S \times \boldsymbol{Y})$ are adequate under our identifiability conditions for learning about the causal quantities of interest.

%%%%%%%%%%%%%%%%%%%%%%%%%%%%%%%%%%%%%%%%%%%%%%%%%%%%%%%%%%%%
\subsection{Augmented inverse probability weighting estimation}
%%%%%%%%%%%%%%%%%%%%%%%%%%%%%%%%%%%%%%%%%%%%%%%%%%%%%%%%%%%%

\paragraph{Influence function of $\psi(a)$ under a nonparametric model $\mathcal M$:} Working at the cluster level, the results in \cite{dahabreh2018generalizing} imply that the influence function of $\psi(a)$ under the nonparametric model $\mathcal M$ is 
\begin{equation*}
    \begin{split}
    \mathit{\Psi}^1_{p_0}(a) &= \dfrac{I(S = 1, A = a)}{\Pr_{p_0}[S = 1 , A  = 1 | X, \boldsymbol{W}]} \Big\{ \overbar{Y} - \E_{p_0}\left[ \overbar{Y} | X, \boldsymbol{W}, S = 1, A = a \right] \Big\} \\
    &\quad\quad\quad\quad\quad\quad\quad\quad+ \E_{p_0}\left[ \overbar{Y} | X, \boldsymbol{W}, S = 1, A = a \right] - \psi_{p_0}(a),
    \end{split}
\end{equation*}
where the subscript $p_0$ indicates the ``true'' data law in the target population; $\Pr_{p_0}[S = 1 , A  = a | X, W] = \Pr_{p_0}[S = 1 | X, \boldsymbol{W}] \Pr_{p_0}[ A  = a | X, \boldsymbol{W}, S = 1]$ is the joint cluster-level probability of participating in the trial and being assigned to treatment $a$, given $(X, \boldsymbol{W})$; and $ \E_{p_0}\left[ \overbar{Y} | X, \boldsymbol{W}, S = 1, A = a \right]$ is the conditional expectation of the cluster-level average observed outcome $ \overbar{Y} $ given $(X, \boldsymbol{W})$ among clusters participating in the trial and assigned to treatment $a$.

\paragraph{Efficient influence function of $\psi(a)$ under a semiparametric model with known probability of treatment in the trial:} We now consider inference about $\psi(a)$ under a semiparametric model, $\mathcal{M}_{\text{\tiny semi}}$ in which the the law of the observed data $(X,\boldsymbol{W}, S,A,\boldsymbol{Y})$ is restricted compared to $\mathcal M$ by assuming that the probability of treatment given covariates and participation status, $p_{A | X, \boldsymbol{W}, S}$, is a known function. Under this semiparametric model, the law for $\boldsymbol{Y}$ given $(X,\boldsymbol{W},S,A)$ is still left unspecified, as is the law for $(X,\boldsymbol{W},S)$. We are interested in $\mathcal{M}_{\text{\tiny semi}}$ because $p_{A | X, \boldsymbol{W}, S}(a|x, \boldsymbol{w}, 1)$ is typically known and under the control of the investigators in the randomized trial, and it is possible that $p_{A | X, \boldsymbol{W}, S}(a|x, \boldsymbol{w}, 0)$ is also known (e.g., if only a single vaccination strategy is available in non-experimental settings). We will argue that the efficient influence function for $\psi(a)$ under this semiparamertic model is the same as the (unique) influence function $ \mathit{\Psi}^1_{p_0}(a)$ under the nonparametric model.

To obtain the \emph{efficient} influence function under the semiparametric model we need to project $ \mathit{\Psi}^1_{p_0}(a)$ onto the tangent space of the semiparametric model; we denote that tangent space as $\Lambda_{\text{\tiny semi}}$. Note that the influence function under the nonparametric model is an influence function under any semiparametric model. 

Note that $\overbar{Y}$ is a function of the vector $\boldsymbol{Y}$, because $$\overbar{Y}_j = \dfrac{\boldsymbol{1}_{1 \times N_j} \boldsymbol{Y}_j^\intercal}{N_j} \equiv u(\boldsymbol{Y}_j),$$ where $\boldsymbol{1}_{1 \times \mbox{\tiny length}(\boldsymbol{Y}_j)}$ is a row vector of 1s with $N_j$ elements. Using the definition above and invoking the law of the unconscious statistician \cite{casella2002statistical}, we can write $\psi(a)$ as 
\begin{equation*}
        \psi(a) = \int \int \int u(\boldsymbol{y}) p_{\boldsymbol{Y}|X,\boldsymbol{W},S,A}(\boldsymbol{y}|x,\boldsymbol{w}, S = 1, A = a) p_{X,\boldsymbol{W}}(x,\boldsymbol{w}) d\boldsymbol{y}  dxd\boldsymbol{w},
\end{equation*}
which shows that $\psi(a)$ does not depend on the law $p_{A|X,\boldsymbol{W},S}$. Thus, $\mathit{\Psi}^1_{p_0}(a)$ is orthogonal to the scores of $p_{A|X,\boldsymbol{W},S}$ and belongs to the orthogonal complement of the scores of $p_{A|X,\boldsymbol{W},S}$, which we denote as $\Lambda_{A|X,\boldsymbol{W},S}^{\bot}$. From the decomposition $L^{0}_2(P) = \Lambda_{\boldsymbol{Y}|X,\boldsymbol{W},S,A} \oplus \Lambda_{A|X,\boldsymbol{W},S} \oplus \Lambda_{S|X,\boldsymbol{W}} \oplus \Lambda_{X,\boldsymbol{W}} $ we have that $\Lambda_{A|X,\boldsymbol{W},S}^{\bot} = \Lambda_{\boldsymbol{Y}|X,\boldsymbol{W},S,A} \oplus \Lambda_{S|X, \boldsymbol{W}} \oplus \Lambda_{X,\boldsymbol{W}}$. Therefore, $\mathit{\Psi}^1_{p_0}(a)$ belongs to $\Lambda_{\boldsymbol{Y}|X,\boldsymbol{W},S,A} \oplus \Lambda_{S|X, \boldsymbol{W}} \oplus \Lambda_{X,\boldsymbol{W}}$. Because $\Lambda_{\text{\tiny semi}}$ is equal to $\Lambda_{\boldsymbol{Y}|X,\boldsymbol{W},S,A} \oplus \Lambda_{S|X, \boldsymbol{W}} \oplus \Lambda_{X,\boldsymbol{W}}$, we conclude that  $\mathit{\Psi}^1_{p_0}(a)$ belongs to $\Lambda_{\text{\tiny semi}}$ and thus its projection onto $\Lambda_{\text{\tiny semi}}$ is equal to the influence function itself \cite{van2000asymptotic}. This implies that the \emph{efficient} influence function for $\psi(a)$ under the semiparametric model where $p_{A | X, \boldsymbol{W}, S}$ is known, is equal to the unique influence function $\mathit{\Psi}^1_{p_0}(a)$ under the nonparametric model.

To verify the above argument, we can explicitly calculate the influence function under the semiparametric model, which we denote $\mathit{\Psi}^1_{\text{\tiny semi}}(a)$. To do so, we first need the projection of $\mathit{\Psi}^1_{p_0}(a)$ onto the orthogonal complement of the tangent space of the semiparametric model, $\prod\left\{ \mathit{\Psi}^1_{p_0}(a) \big| \Lambda_{\text{\tiny semi}}^\bot \right\} = \prod\left\{ \mathit{\Psi}^1_{p_0}(a) \big| \Lambda_{A|X,\boldsymbol{W},S} \right\}$. From Theorem 4.5 of \cite{tsiatis2007} and standard conditional expectation arguments we obtain $\prod\left\{ \mathit{\Psi}^1_{p_0}(a) \big| \Lambda_{A|X,\boldsymbol{W},S} \right\} = \E\left[ \mathit{\Psi}^1_{p_0}(a) \big| X, \boldsymbol{W}, S, A \right] - \E\left[\mathit{\Psi}^1_{p_0}(a) | X, \boldsymbol{W}, S \right] = 0$. This result confirms the argument above, because by Theorem 4.3 of \cite{tsiatis2007} the efficient influence function under the semiparametric model is $\mathit{\Psi}^1_{\text{\tiny semi}}(a) = \prod\left\{ \mathit{\Psi}^1_{p_0}(a) \big| \Lambda_{\text{\tiny semi}}\right\} = \mathit{\Psi}^1_{p_0}(a) - \prod\left\{ \mathit{\Psi}^1_{p_0}(a) \big| \Lambda_{\text{\tiny semi}}^\bot \right\} = \mathit{\Psi}^1_{p_0}(a).$

\paragraph{One-step in-sample estimator of $\psi(a)$:} The influence function $\mathit{\Psi}^1_{p_0}(a)$ suggests the augmented inverse probability weighting estimator,
\begin{equation}
    \begin{split}
    \widehat{\psi}(a) &= \dfrac{1}{m} \sum\limits_{j = 1}^{m} \Bigg\{ \dfrac{I(S_j = 1, A_j = a)}{\widehat p(X_j, \boldsymbol{W}_j) \widehat e_a(X_j, \boldsymbol{W}_j)}  \Big\{ \overbar{Y}_j - \widehat g_a(X_j, \boldsymbol{W}_j) \Big\} + \widehat g_a(X_j, \boldsymbol{W}_j) \Bigg\}.
    \end{split}
\end{equation}
We refer to this estimator as a one-step in-sample estimator because to obtain it we use all the available data to estimate $\widehat p(X, \boldsymbol{W})$, $\widehat e_a(X, \boldsymbol{W})$, and $\widehat g_a(X, \boldsymbol{W})$, and then use the same data to calculate the value of $\widehat \psi(a)$.

%%%%%%%%%%%%%%%%%%%%%%%%%%%%%%%%%%%%%%%%%%%%%%%%%%%%%%%%%%%%
\subsection{Asymptotic behavior of \texorpdfstring{$\widehat{\psi}(a)$}{e}}
%%%%%%%%%%%%%%%%%%%%%%%%%%%%%%%%%%%%%%%%%%%%%%%%%%%%%%%%%%%%

We now give, without proof, a result about the asymptotic distribution of $\widehat{\psi}(a)$, in the general case when both $\Pr[S = 1 |  X, \boldsymbol{W} ]$ and $\E [\overbar{Y} |  X, \boldsymbol{W}, S = 1, A = a]$ are unknown and need to be estimated. A proof of this result can be obtained by following the same steps as our previous work on generalizability analyses with individually randomized trials (e.g., \cite{dahabreh2019generalizing}).

For arbitrary functions $p'(X, \boldsymbol{W})$, $e'_a(X, \boldsymbol{W})$, and $g'_a(X, \boldsymbol{W})$ define $$H\big( p'(X, \boldsymbol{W}), e'_a(X, \boldsymbol{W}), g'_a(X, \boldsymbol{W}) \big) = \dfrac{I(S=1, A = a)}{p'(X,\boldsymbol{W}) e'_a(X, \boldsymbol{W}) } \big\{ \overbar{Y} - g'_a(X, \boldsymbol{W}) \big\} + g'_a(X, \boldsymbol{W}).$$ Furthermore, following \cite{van1996weak}, define $\mathbb{P}_m\big(v(W)\big) = \frac{1}{m} \sum_{j=1}^m v(W_j)$ and $\mathbb{G}_m(v(W)) = \sqrt{m}\big(\mathbb{P}_m(v(W)) - \E[v(W)]\big)$, for some function $v$ and a random variable $W$. Note that $$\widehat \psi(a) = \mathbb{P}_m \Big( H \big( \widehat p(X, \boldsymbol{W}),  \widehat e_a(X, \boldsymbol{W}),  \widehat g_a(X, \boldsymbol{W}) \big) \Big).$$ 

Throughout, we assume that $\widehat p(X, \boldsymbol{W})$, $\widehat e_a(X, \boldsymbol{W})$, and $\widehat g_a(X, \boldsymbol{W})$ have well-defined limits as $m \rightarrow \infty$, which we denote as $p^*(X, \boldsymbol{W})$, $e_a^*(X, \boldsymbol{W})$, and $g_a^*(X, \boldsymbol{W})$, respectively. To derive the asymptotic representation for $\widehat{\psi}(a)$ we make the following assumptions: 
\begin{enumerate}
    \item[(i)] $H \big( \widehat p(X, \boldsymbol{W}),  \widehat e_a(X, \boldsymbol{W}),  \widehat g_a(X, \boldsymbol{W}) \big)$ and the limit $H \big( p^*(X, \boldsymbol{W}),  e^*_a(X, \boldsymbol{W}), g^*_a(X, \boldsymbol{W}) \big)$ belong to a Donsker class; 
    \item[(ii)] $\Big|\Big| H \big( \widehat p(X, \boldsymbol{W}),  \widehat e_a(X, \boldsymbol{W}),  \widehat g_a(X, \boldsymbol{W}) \big)  - H \big( p^*(X, \boldsymbol{W}),  e^*_a(X, \boldsymbol{W}), g^*_a(X, \boldsymbol{W}) \big) \Big|\Big|_2 \overset{p}{\longrightarrow} 0$;
    \item[(iii)] $\E\left[ \Big( H \big( p^*(X, \boldsymbol{W}),  e^*_a(X, \boldsymbol{W}), g^*_a(X, \boldsymbol{W}) \big) \Big)^2 \right] < 0$;
    \item[(iv)] At least one of the following two assumptions holds: $\widehat p(X, \boldsymbol{W}) \overset{p}{\longrightarrow} \Pr[S = 1 | X, \boldsymbol{W} ]$ or $\widehat g_a(X, \boldsymbol{W}) \overset{p}{\longrightarrow} \E[ \overbar{Y} | X, \boldsymbol{W} , S = 1, A =a ]$;
    \item[(v)] $\widehat e_a(X,  \boldsymbol{W})$ is $\sqrt{m}$-consistent for $\Pr[A = a |  X, \boldsymbol{W} , S = 1]$.
\end{enumerate}
Note, in passing, that the Donsker conditions in assumption (i) can be relaxed by modifying $\widehat \psi(a)$ using the approach described in \cite{chernozhukov2018double}.

We now give a general result for the asymptotic properties of $\widehat \psi(a)$.
\begin{proposition}
If assumptions \emph{(i)} through \emph{(v)} hold, then, as $m \longrightarrow \infty$,
\begin{enumerate}
    \item $\widehat \psi (a) \overset{p}{\longrightarrow} \psi(a)$; and
    \item $\widehat \psi(a)$ has the following asymptotic representation, 
    \begin{equation*} 
        \begin{split} 
        &\sqrt{m} \big( \widehat \psi (a) - \psi (a) \big) =  \mathbb{G}_m  \Big( H \big( p^*(X, \boldsymbol{W}),  e^*_a(X, \boldsymbol{W}),  g^*_a(X, \boldsymbol{W}) \big) \Big)   + R + o_P(1),
    \end{split}    
     \end{equation*}
     where $$R \leq \sqrt{m} O_P\left( \Big|\Big| \widehat p(X, \boldsymbol{W}) -  \Pr[S = 1 | X, \boldsymbol{W} ] \Big|\Big|_2  \Big|\Big|  \widehat g_a(X, \boldsymbol{W}) -   \E[ \overbar{Y} | X, \boldsymbol{W} , S = 1, A =a ]   \Big|\Big|_2 \right).$$
\end{enumerate}
\end{proposition}
In the asymptotic representation above, the first term on the right-hand-side is asymptotically normal. Thus, the asymptotic behavior is driven by the second term. 

When the second term is $o_P(1)$ (i.e., when both $ \widehat p(X, \boldsymbol{W})$ and $\widehat g_a(X, \boldsymbol{W})$ are consistent and converge at sufficiently fast rate to the corresponding true conditional functions), then $\widehat \psi(a)$ is consistent, asymptotically normal, and semiparametric efficient. That is to say, the asymptotic distribution of $\sqrt{m}\big( \widehat \psi(a) - \psi(a) \big)$ is normal with mean zero and variance 
\begin{equation}\label{eq:as_var_DR}
    \begin{split}
    \sigma^2_{\widehat \psi(a)} &= \E_{p_0} \left[   \Big( \mathit{\Psi}^1_{p_0}(a) \Big)^2 \right] \\
    &= \E_{p_0} \left[ \dfrac{\mbox{Var}_{p_0} [\overbar{Y} |  X, \boldsymbol{W} , S = 1, A = a ] }{\Pr_{p_0}[S = 1, A = a |  X, \boldsymbol{W} ]} + \big( \E_{p_0}[\overbar{Y} |X, \boldsymbol{W}, S = 1, A = a ] \big)^{2} - \big( \psi_{p_0}(a) \big)^2 \right].
    \end{split}
\end{equation}

When the second term is just $O_P(1)$, $\widehat \psi(a)$ remains consistent and asymptotically normal (the double robustness property), but is no longer efficient.

%%%%%%%%%%%%%%%%%%%%%%%%%%%%%%%%%%%%%%%%%%%%%%%%%%%%%%%%%%%%
\subsection{Inverse probability weighting and g-formula estimators}
%%%%%%%%%%%%%%%%%%%%%%%%%%%%%%%%%%%%%%%%%%%%%%%%%%%%%%%%%%%%

The augmented estimator in the main text can be modified to obtain a non-augmented inverse probability weighting estimator, 
\begin{equation}\label{eq:estimator_IP_weighting}
    \begin{split}
    \widehat{\psi}_{\text{\tiny w}}(a) &= \dfrac{1}{m} \sum\limits_{j = 1}^{m} \dfrac{I(S_j = 1, A_j = a) \overbar{Y}_j}{\widehat p(X_j, \boldsymbol{W}_j) \widehat e_a(X_j, \boldsymbol{W}_j)}.
    \end{split}
\end{equation}
Of note, $\widehat{\psi}_{\text{\tiny w}}(a)$ is consistent only when $\widehat p(X, \boldsymbol{W})$ is a consistent estimator of $\Pr[S = 1 | X , \boldsymbol{W}]$; $\widehat e_a(X, \boldsymbol{W})$ can always be chosen to be consistent because the randomization probabilities are known.

Usually, the version of the weighted estimator that normalizes the weights by their sum will perform better, 
\begin{equation}\label{eq:estimator_IP_weightingHajek}
    \begin{split}
    \widetilde{\psi}_{\text{\tiny w}}(a) &= \left \{ \sum\limits_{j = 1}^{m} \dfrac{I(S_j = 1, A_j = a) }{\widehat p(X_j, \boldsymbol{W}_j) \widehat e_a(X_j, \boldsymbol{W}_j)} \right\}^{-1} \sum\limits_{j = 1}^{m} \dfrac{I(S_j = 1, A_j = a) \overbar{Y}_j}{\widehat p(X_j, \boldsymbol{W}_j) \widehat e_a(X_j, \boldsymbol{W}_j)}.
    \end{split}
\end{equation}

These inverse probability weighting estimators are closely related to the weighting estimators discussed in prior work on generalizability with cluster randomized trials \cite{tipton2012, tipton2013improving, tipton2014}. If we assume that only cluster-level variables (including any aggregates of individual-level covariates), $X$, are adequate to estimate the probability of trial participation and the probability of treatment among randomized clusters, and we obtain cluster level assessments of the outcome, $\overbar{Y}$, then we can write the estimator as
\begin{equation}\label{eq:estimator_IP_weighting_agg}
    \begin{split}
    \widehat{\psi}^{\text{\tiny agg}}_{\text{\tiny w}}(a) &= \dfrac{1}{m} \sum\limits_{j = 1}^{m} \dfrac{I(S_j = 1, A_j = a) \overbar{Y}_j}{\widehat p(X_j) \widehat e_a(X_j)},
    \end{split}
\end{equation}
highlighting that it only uses aggregated cluster-level data or estimates that can be obtained from such data, but does not require any individual-level data.

Alternatively, we might consider the g-formula estimator, 
\begin{equation}\label{eq:estimator_g_form}
   \widehat{\psi}_{\text{\tiny g}}(a)  = \dfrac{1}{m} \sum\limits_{j = 1}^{m}  \widehat g_a(X_j, \boldsymbol{W}_j),
\end{equation}
which is consistent only when $ \widehat g_a(X, \boldsymbol{W})$ is a consistent estimator of $\E[\overbar{Y}|X,\boldsymbol{W}, S = 1, A = a]$ (correct specification of a model for this conditional expectation function is unlikely because of complex within-cluster dependence).

The results in\cite{hahn1998role} and \cite{tan2007} can be used to argue that if all three estimators -- $\widehat p(X, \boldsymbol{W})$, $\widehat g_a(X, \boldsymbol{W})$, and $\widehat e_a(X_j, \boldsymbol{W}_j)$  -- are consistent and converge at sufficiently fast rate, then $\widehat{\psi}(a)$ is semiparametric efficient and has asymptotic variance smaller than or equal to the asymptotic variance of $\widehat{\psi}_{\text{\tiny w}}(a)$, but larger than or equal to the asymptotic variance of $\widehat{\psi}_{\text{\tiny g}}(a)$. The practical importance of the difference between $\widehat{\psi}(a)$ and $\widehat{\psi}_{\text{\tiny g}}(a)$ is debatable because correct specification of the model for the outcome is extremely challenging in the presence of interference and in practice $\widehat g_a(X, \boldsymbol{W})$ will often be inconsistent. Perhaps more important for practice is that experience with similar estimators with individually randomized trials, as well as simulation studies \cite{balzer2019new} for the analysis of cluster randomized trials (not in a generalizability context), suggest that when the model for $\widehat p(X, \boldsymbol{W})$ is correctly specified so that both $\widehat{\psi}(a)$ and $\widehat{\psi}_{\text{\tiny w}}(a)$ are consistent, it is usually (but not always \cite{kang2007}) the case that the former is more efficient than the latter.

%%%%%%%%%%%%%%%%%%%%%%%%%%%%%%%%%%%%%%%%%%%%%%%%%%%%%%%%%%%%
%%%%%%%%%%%%%%%%%%%%%%%%%%%%%%%%%%%%%%%%%%%%%%%%%%%%%%%%%%%%
\clearpage
\section{Identification and estimation of \texorpdfstring{$\E\left[\overbar{Y}^{a} | S = 0 \right]$}{e}}
\renewcommand{\theequation}{B.\arabic{equation}}
\setcounter{equation}{0}
%%%%%%%%%%%%%%%%%%%%%%%%%%%%%%%%%%%%%%%%%%%%%%%%%%%%%%%%%%%%
%%%%%%%%%%%%%%%%%%%%%%%%%%%%%%%%%%%%%%%%%%%%%%%%%%%%%%%%%%%%

In this section, we summarize identification and estimation results for potential outcome means in the non-randomized subset of the target population, that is  to say, $\E[\overbar{Y}^a | S = 0]$, when using data from the same nested trial design described in the main text.

\paragraph{Identifiability conditions:} Consider the following slight modification of identifiability condition \emph{A5} from the main text: 

\vspace{0.1in}
\noindent
\emph{B5. Positivity of trial participation:} $\Pr[S = 1 | X = x, \boldsymbol{W} = \boldsymbol{w}] > 0$ for every $x$ and $\boldsymbol{w}$ with positive density among non-randomized individuals in the target population.

\paragraph{Identification:} We now state an identification result for $\E[\overbar{Y}^a | S = 0]$.

\begin{proposition}
Under identifiability conditions A1 through A4 from the main text and condition B5, the expectation of the average potential outcome in the target population, $ \E\left[\overbar{Y}^{a} | S = 0 \right] $, is identified by $$\phi(a) \equiv \E \Big[ \E\left[\overbar{Y} | X, \boldsymbol{W}, S = 1, A = a \right] \big| S = 0 \Big].$$ 
\end{proposition}

Starting with the causal quantity of interest,
\begin{equation*}
    \begin{split}
        \E\left[\overbar{Y}^{a} | S = 0 \right] &= \E \Big[ \E\left[\overbar{Y}^{a} | X, \boldsymbol{W}, S = 0 \right] \big| S = 0  \Big] \\
            &= \E \Big[ \E\left[\overbar{Y}^{a} | X, \boldsymbol{W}, S = 1 \right]  \big| S = 0  \Big] \\
            &= \E \Big[ \E\left[\overbar{Y}^{a} | X, \boldsymbol{W}, S = 1, A = a \right]  \big| S = 0  \Big] \\
            &= \E \Big[ \E\left[\overbar{Y} | X, \boldsymbol{W}, S = 1, A = a \right]  \big| S = 0  \Big] \\
            &= \phi(a),
    \end{split}
\end{equation*}
where the first step follows from the law of total expectation, the second by condition \emph{A4}, the third by condition \emph{A2}, the fourth by condition \emph{A1}, the last by the definition of $\phi(a)$; quantities are well-defined because of the positivity conditions \emph{A3} and \emph{B5}. We note that identification of $\E[\overbar{Y}^a | S = 0]$ is also possible under weaker conditions of exchangeability in mean rather than in distribution (e.g., see \cite{dahabreh2019transportingStatMed} for arguments in the individually-randomized trial case that can be adapted to the cluster randomized trial case).

\paragraph{Identification of the average treatment effect among the non-randomized population under weaker conditions:} Following earlier work \cite{dahabreh2019transportingStatMed}, if the causal quantity of interest is only the average treatment effect among the non-randomized individuals (but not the treatment-specific potential outcome means), then identification is possible under weaker conditions. Specifically, consider the following identification condition: 

\vspace{0.1in}
\noindent
\emph{B4. Conditional exchangeability in measure over $S$:} for treatments $a \in \mathcal A$ and $a^\prime \in \mathcal A$, $\E[\overbar{Y}^a - \overbar{Y}^{a^\prime} | X = x, \boldsymbol{W} = \boldsymbol{w} , S = 0 ] = \E[\overbar{Y}^a - \overbar{Y}^{a^\prime} | X = x, \boldsymbol{W} = \boldsymbol{w}, S = 1]$ for all $x$ and $\boldsymbol{w}$ with positive density among non-randomized individuals in the target population.

\begin{proposition}
Under conditions A1 through A3 from the main text, B4, and B5, the average treatment effect in the non-randomize subset of the target population, $ \E\Big[\overbar{Y}^{a} - \overbar{Y}^{a^\prime} \big| S = 0 \Big]$, is identifiable by $$\phi(a, a^\prime) \equiv \E \Big[ \E\Big[\overbar{Y} | X, \boldsymbol{W}, S = 1, A = a \Big] - \E\Big[\overbar{Y} | X, \boldsymbol{W}, S = 1, A = a^\prime \Big] \big| S = 0  \Big].$$

\end{proposition}

Starting with the causal quantity of interest,
\begin{equation*}
    \begin{split}
        \E\Big[\overbar{Y}^{a} - \overbar{Y}^{a^\prime} \big| S = 0 \Big] &= \E \Big[ \E\left[\overbar{Y}^{a} -  \overbar{Y}^{a^\prime}| X, \boldsymbol{W} , S = 0 \right] \big| S = 0  \Big] \\
            &= \E \Big[ \E\left[\overbar{Y}^{a} - \overbar{Y}^{a^\prime} | X, \boldsymbol{W}, S = 1 \right] \big| S = 0  \Big] \\
            &= \E \Big[ \E\Big[\overbar{Y}^{a} | X, \boldsymbol{W}, S = 1, A = a \Big] - \E\Big[\overbar{Y}^{a^\prime} | X, \boldsymbol{W}, S = 1, A = a^\prime \Big] \big| S = 0  \Big] \\
            &= \E \Big[ \E\Big[\overbar{Y} | X, \boldsymbol{W}, S = 1, A = a \Big] - \E\Big[\overbar{Y} | X, \boldsymbol{W}, S = 1, A = a^\prime \Big] \big| S = 0  \Big] \\
            &= \phi(a, a^\prime),
    \end{split}
\end{equation*}
where the first step follows from the law of total expectation, the second by condition \emph{B4}, the third by condition \emph{A2}, the fourth by condition \emph{A1}, and the last by the definition of $\phi(a, a^\prime)$; quantities are well-defined because of the positivity conditions \emph{A3} and \emph{B5}.

\paragraph{Augmented inverse odds weighting estimator:} Working at the cluster level, the results in \cite{dahabreh2019transportingStatMed} imply that the influence function of $\phi(a)$ under the nonparametric model is 
\begin{equation*}
    \begin{split}
    \mathit{\Phi}^1_{p_0}(a) &= \dfrac{1}{\Pr_{p_0}[S = 0 ]} \Bigg\{ \dfrac{I(S = 1, A = a) \Pr_{p_0}[S = 0| X, \boldsymbol{W}] }{\Pr_{p_0}[S = 1 , A  = 1 | X, \boldsymbol{W}]} \Big\{ \overbar{Y} - \E_{p_0}\left[ \overbar{Y} | X, \boldsymbol{W}, S = 1, A = a \right] \Big\} \\
    &\quad\quad\quad\quad\quad\quad\quad\quad+ I(S = 0) \Big\{ \E_{p_0}\left[ \overbar{Y} | X, \boldsymbol{W}, S = 1, A = a \right] - \phi_{p_0}(a) \Big\} \Bigg\},
    \end{split}
\end{equation*}
where the subscript $p_0$ indicates the ``true'' data law in the target population; $\Pr_{p_0}[S = 1 , A  = a | X, W] = \Pr_{p_0}[S = 1 | X, \boldsymbol{W}] \Pr_{p_0}[ A  = a | X, \boldsymbol{W}, S = 1]$ is the joint cluster-level probability of participating in the trial and being assigned to treatment $a$, given $(X, \boldsymbol{W})$; $\Pr_{p_0}[S = 0| X, \boldsymbol{W}]$ is the probability of not participating in the trial given $(X, \boldsymbol{W})$ ; and $ \E_{p_0}\left[ \overbar{Y} | X, \boldsymbol{W}, S = 1, A = a \right]$ is the conditional expectation of the cluster-level average observed outcome $ \overbar{Y} $ given $(X, \boldsymbol{W})$ among clusters participating in the trial and assigned to treatment $a$.

As similar argument as for $\psi(a)$ (see Appendix \ref{appendix_A}) can be used to show that the efficient influence function for $\phi(a)$ under a semiparameric model where $p_{A | X, \boldsymbol{W}, S}$ is known is the same as the unique influence function under the nonparametric model $\mathit{\Phi}^1_{p_0}(a)$, given above.

The influence function $\mathit{\Phi}^1_{p_0}(a)$ suggests the augmented inverse odds weighting estimator
\begin{equation*}
    \begin{split}
    \widehat{\phi}(a) &= \left\{\sum\limits_{j = 1}^{m} I(S_j = 0)\right\}^{-1} \sum\limits_{j = 1}^{m} \Bigg\{  \widehat w_a(X_j, \boldsymbol{W}_j, S_j, A_j ) \Big\{ \overbar{Y}_j -  \widehat g_a(X_j, \boldsymbol{W}_j) \Big\} + I(S_j = 0) \widehat g_a(X_j, \boldsymbol{W}_j) \Bigg\} ,
    \end{split}
\end{equation*}
with $$ \widehat w_a(X_j, \boldsymbol{W}_j, S_j, A_j ) =  \dfrac{I(S_j = 1, A_j = a) \big[ 1- \widehat p(X_j, \boldsymbol{W}_j) \big]}{\widehat p(X_j, \boldsymbol{W}_j) \widehat e_a(X_j, \boldsymbol{W}_j)}.$$

We end by noting that the identification and estimation results in this section of the Appendix also apply, with appropriate modifications, to data obtained from the non-nested trial design \cite{dahabreh2019studydesigns,  dahabreh2019transportingStatMed}, where randomized and non-randomized clusters are sampled separately.

%%%%%%%%%%%%%%%%%%%%%%%%%%%%%%%%%%%%%%%%%%%%%%%%%%%%%%%%%%%%%%%%%%%%%%%%%%%%%%
% BIBLIOGRAPHY
%%%%%%%%%%%%%%%%%%%%%%%%%%%%%%%%%%%%%%%%%%%%%%%%%%%%%%%%%%%%%%%%%%%%%%%%%%%%%%
\clearpage
\renewcommand{\refname}{Appendix References}
\bibliographystyle{unsrt}
%\bibliography{bibliography}

\begin{thebibliography}{10}

\bibitem{donner1994cluster}
Allan Donner and Neil Klar.
\newblock Cluster randomization trials in epidemiology: theory and application.
\newblock {\em Journal of Statistical Planning and Inference}, 42(1-2):37--56,
  1994.

\bibitem{hernan2016discussionkeiding}
M~A Hern{\'a}n.
\newblock Discussion of ``{P}erils and potentials of self-selected entry to
  epidemiological studies and surveys''.
\newblock {\em Journal of the Royal Statistical Society. Series A (Statistics
  in Society)}, 179(2):346--347, 2016.

\bibitem{dahabreh2019commentaryonweiss}
Issa~J Dahabreh and Miguel~A Hern{\'a}n.
\newblock Extending inferences from a randomized trial to a target population.
\newblock {\em European Journal of Epidemiology}, pages 1--4, 2019.

\bibitem{dahabreh2018generalizing}
Issa~J Dahabreh, Sarah~E Robertson, Eric~J Tchetgen~Tchetgen, Elizabeth~A
  Stuart, and Miguel~A Hern{\'a}n.
\newblock Generalizing causal inferences from individuals in randomized trials
  to all trial-eligible individuals.
\newblock {\em Biometrics}, 75(2):685--694, 2018.

\bibitem{buchanan2018generalizing}
Ashley~L Buchanan, Michael~G Hudgens, Stephen~R Cole, Katie~R Mollan, Paul~E
  Sax, Eric~S Daar, Adaora~A Adimora, Joseph~J Eron, and Michael~J Mugavero.
\newblock Generalizing evidence from randomized trials using inverse
  probability of sampling weights.
\newblock {\em Journal of the Royal Statistical Society. Series A (Statistics
  in Society)}, 181(4):1193--1209, 2018.

\bibitem{lesko2017practical}
Catherine~R Lesko, Ashley~L Buchanan, Daniel Westreich, Jessie~K Edwards,
  Michael~G Hudgens, and Stephen~R Cole.
\newblock Generalizing study results: A potential outcomes perspective.
\newblock {\em Epidemiology}, 28(4):553--561, 2017.

\bibitem{balzer2019new}
Laura~B Balzer, Wenjing Zheng, Mark~J van~der Laan, and Maya~L Petersen.
\newblock A new approach to hierarchical data analysis: Targeted maximum
  likelihood estimation for the causal effect of a cluster-level exposure.
\newblock {\em Statistical Methods in Medical Research}, 28(6):1761--1780,
  2019.

\bibitem{benitez2022comparative}
Alejandra Benitez, Maya~L Petersen, Mark~J van~der Laan, Nicole Santos,
  Elizabeth Butrick, Dilys Walker, Rakesh Ghosh, Phelgona Otieno, Peter Waiswa,
  and Laura~B Balzer.
\newblock Defining and estimating effects in cluster randomized trials: A
  methods comparison.
\newblock {\em arXiv preprint arXiv:2110.09633}, 2022.

\bibitem{omuircheartaigh2014}
Colm O'Muircheartaigh and Larry~V Hedges.
\newblock Generalizing from unrepresentative experiments: a stratified
  propensity score approach.
\newblock {\em Journal of the Royal Statistical Society. Series C (Applied
  Statistics)}, 63(2):195--210, 2014.

\bibitem{tipton2013improving}
Elizabeth Tipton.
\newblock Improving generalizations from experiments using propensity score
  subclassification: Assumptions, properties, and contexts.
\newblock {\em Journal of Educational and Behavioral Statistics},
  38(3):239--266, 2013.

\bibitem{tipton2017implications}
Elizabeth Tipton, Kelly Hallberg, Larry~V Hedges, and Wendy Chan.
\newblock Implications of small samples for generalization: Adjustments and
  rules of thumb.
\newblock {\em Evaluation Review}, 41(5):472--505, 2017.

\bibitem{dahabreh2019studydesigns}
Issa~J Dahabreh, Sebastien J-PA Haneuse, James~M Robins, Sarah~E Robertson,
  Ashley~L Buchanan, Elisabeth~A Stuart, and Miguel~A Hern\'an.
\newblock Study designs for extending causal inferences from a randomized trial
  to a target population.
\newblock {\em arXiv preprint arXiv:1905.07764}, 2019.

\bibitem{rubin1974}
Donald~B Rubin.
\newblock Estimating causal effects of treatments in randomized and
  nonrandomized studies.
\newblock {\em Journal of {E}ducational {P}sychology}, 66(5):688, 1974.

\bibitem{robins2000d}
James~M Robins and Sander Greenland.
\newblock Causal inference without counterfactuals: comment.
\newblock {\em Journal of the American Statistical Association},
  95(450):431--435, 2000.

\bibitem{pearl2014}
Judea Pearl and Elias Bareinboim.
\newblock External validity: from do-calculus to transportability across
  populations.
\newblock {\em Statistical Science}, 29(4):579--595, 2014.

\bibitem{bareinboim2016causalfusion}
Elias Bareinboim and Judea Pearl.
\newblock Causal inference and the data-fusion problem.
\newblock {\em Proceedings of the National Academy of Sciences},
  113(27):7345--7352, 2016.

\bibitem{richardson2013single}
Thomas~S Richardson and James~M Robins.
\newblock Single world intervention graphs ({S}{W}{I}{G}s): A unification of
  the counterfactual and graphical approaches to causality.
\newblock Technical Report 128, Center for Statistics and the Social Sciences,
  University of Washington, 2013.

\bibitem{dahabreh2019identification}
Issa~J Dahabreh, James~M Robins, Sebastien J-PA Haneuse, and Miguel~A Hern\'an.
\newblock Generalizing causal inferences from randomized trials: counterfactual
  and graphical identification.
\newblock {\em arXiv preprint arXiv:1906.10792}, 2019.

\bibitem{vanderWeele2012}
Tyler~J VanderWeele.
\newblock Confounding and effect modification: distribution and measure.
\newblock {\em Epidemiologic Methods}, 1(1):55--82, 2012.

\bibitem{efron1994introduction}
Bradley Efron and Robert~J Tibshirani.
\newblock {\em An introduction to the bootstrap}, volume~57 of {\em Monographs
  on Statistics and Applied Probability}.
\newblock Chapman \& Hall/CRC, 1994.

\bibitem{gravenstein2016cluster}
Stefan Gravenstein, Roshani Dahal, Pedro~L Gozalo, H~Edward Davidson, Lisa~F
  Han, Monica Taljaard, and Vincent Mor.
\newblock A cluster randomized controlled trial comparing relative
  effectiveness of two licensed influenza vaccines in us nursing homes: Design
  and rationale.
\newblock {\em Clinical Trials}, 13(3):264--274, 2016.

\bibitem{gravenstein2017comparative}
Stefan Gravenstein, H~Edward Davidson, Monica Taljaard, Jessica Ogarek, Pedro
  Gozalo, Lisa Han, and Vincent Mor.
\newblock Comparative effectiveness of high-dose versus standard-dose influenza
  vaccination on numbers of us nursing home residents admitted to hospital: a
  cluster-randomised trial.
\newblock {\em The Lancet Respiratory Medicine}, 5(9):738--746, 2017.

\bibitem{hastie1990monographs}
Trevor~J Hastie and Robert~J Tibshirani.
\newblock {\em Generalized additive models}, volume~43 of {\em Monographs on
  statistics and applied probability}.
\newblock Chapman and Hall, 1990.

\bibitem{tibshirani1996regression}
Robert Tibshirani.
\newblock Regression shrinkage and selection via the lasso.
\newblock {\em Journal of the Royal Statistical Society: Series B
  (Methodological)}, 58(1):267--288, 1996.

\bibitem{zou2005regularization}
Hui Zou and Trevor Hastie.
\newblock Regularization and variable selection via the elastic net.
\newblock {\em Journal of the Royal Statistical Society: Series B (Statistical
  Methodology)}, 67(2):301--320, 2005.

\bibitem{breiman2001random}
Leo Breiman.
\newblock Random forests.
\newblock {\em Machine Learning}, 45(1):5--32, 2001.

\bibitem{dahabreh2019relation}
Issa~J Dahabreh, Sarah~E Robertson, and Miguel~A Hern{\'a}n.
\newblock On the relation between g-formula and inverse probability weighting
  estimators for generalizing trial results.
\newblock {\em Epidemiology}, 30(6):807--812, 2019.

\bibitem{chernozhukov2018double}
Victor Chernozhukov, Denis Chetverikov, Mert Demirer, Esther Duflo, Christian
  Hansen, Whitney Newey, and James Robins.
\newblock Double/debiased machine learning for treatment and structural
  parameters.
\newblock {\em The Econometrics Journal}, 21(1):C1--C68, 2018.

\bibitem{cameron2015practitioner}
A~Colin Cameron and Douglas~L Miller.
\newblock A practitioner’s guide to cluster-robust inference.
\newblock {\em Journal of Human Resources}, 50(2):317--372, 2015.

\bibitem{halloran1995causal}
M~Elizabeth Halloran and Claudio~J Struchiner.
\newblock Causal inference in infectious diseases.
\newblock {\em Epidemiology}, pages 142--151, 1995.

\bibitem{halloran2016dependent}
M~Elizabeth Halloran and Michael~G Hudgens.
\newblock Dependent happenings: a recent methodological review.
\newblock {\em Current epidemiology reports}, 3(4):297--305, 2016.

\bibitem{bang2005}
Heejung Bang and James~M Robins.
\newblock Doubly robust estimation in missing data and causal inference models.
\newblock {\em Biometrics}, 61(4):962--973, 2005.

\end{thebibliography}


\begin{thebibliography}{10}

\bibitem{dahabreh2018generalizing}
Issa~J Dahabreh, Sarah~E Robertson, Eric~J Tchetgen~Tchetgen, Elizabeth~A
  Stuart, and Miguel~A Hern{\'a}n.
\newblock Generalizing causal inferences from individuals in randomized trials
  to all trial-eligible individuals.
\newblock {\em Biometrics}, 75(2):685--694, 2018.

\bibitem{casella2002statistical}
George Casella and Roger~L Berger.
\newblock {\em Statistical inference}, volume~2.
\newblock Duxbury Pacific Grove, CA, 2002.

\bibitem{van2000asymptotic}
Aad~W Van~der Vaart.
\newblock {\em Asymptotic statistics}, volume~3.
\newblock Cambridge University Press, 2000.

\bibitem{tsiatis2007}
Anastasios Tsiatis.
\newblock {\em Semiparametric theory and missing data}.
\newblock Springer Science \& Business Media, 2007.

\bibitem{dahabreh2019generalizing}
Issa~J Dahabreh, Miguel~A Hern{\'a}n, Sarah~E Robertson, Ashley Buchanan, and
  Jon~A Steingrimsson.
\newblock Generalizing trial findings in nested trial designs with sub-sampling
  of non-randomized individuals.
\newblock {\em arXiv preprint arXiv:1902.06080}, 2019.

\bibitem{van1996weak}
Aad~W van~der Vaart and Jon~A Wellner.
\newblock {\em Weak Convergence and Empirical Processes}.
\newblock Springer, 1996.

\bibitem{chernozhukov2018double}
Victor Chernozhukov, Denis Chetverikov, Mert Demirer, Esther Duflo, Christian
  Hansen, Whitney Newey, and James Robins.
\newblock Double/debiased machine learning for treatment and structural
  parameters.
\newblock {\em The Econometrics Journal}, 21(1):C1--C68, 2018.

\bibitem{tipton2012}
Elizabeth Tipton.
\newblock Improving generalizations from experiments using propensity score
  subclassification assumptions, properties, and contexts.
\newblock {\em Journal of Educational and Behavioral Statistics},
  38(3):239--266, 2012.

\bibitem{tipton2013improving}
Elizabeth Tipton.
\newblock Improving generalizations from experiments using propensity score
  subclassification: Assumptions, properties, and contexts.
\newblock {\em Journal of Educational and Behavioral Statistics},
  38(3):239--266, 2013.

\bibitem{tipton2014}
Elizabeth Tipton, Larry Hedges, Michael Vaden-Kiernan, Geoffrey Borman, Kate
  Sullivan, and Sarah Caverly.
\newblock Sample selection in randomized experiments: A new method using
  propensity score stratified sampling.
\newblock {\em Journal of Research on Educational Effectiveness},
  7(1):114--135, 2014.

\bibitem{hahn1998role}
Jinyong Hahn.
\newblock On the role of the propensity score in efficient semiparametric
  estimation of average treatment effects.
\newblock {\em Econometrica}, 66(2):315--331, 1998.

\bibitem{tan2007}
Zhiqiang Tan.
\newblock Comment: Understanding {O}{R}, {P}{S} and {D}{R}.
\newblock {\em Statistical {S}cience}, 22(4):560--568, 2007.

\bibitem{balzer2019new}
Laura~B Balzer, Wenjing Zheng, Mark~J van~der Laan, and Maya~L Petersen.
\newblock A new approach to hierarchical data analysis: Targeted maximum
  likelihood estimation for the causal effect of a cluster-level exposure.
\newblock {\em Statistical Methods in Medical Research}, 28(6):1761--1780,
  2019.

\bibitem{kang2007}
Joseph~DY Kang and Joseph~L Schafer.
\newblock Demystifying double robustness: A comparison of alternative
  strategies for estimating a population mean from incomplete data.
\newblock {\em Statistical {S}cience}, pages 523--539, 2007.

\bibitem{dahabreh2019transportingStatMed}
Issa~J Dahabreh, Sarah~E Robertson, Jon~A Steingrimsson, Elizabeth~A Stuart,
  and Miguel~A Hern{\'a}n.
\newblock Extending inferences from a randomized trial to a new target
  population.
\newblock {\em arXiv preprint arXiv:1805.00550}, 2019.

\bibitem{dahabreh2019studydesigns}
Issa~J Dahabreh, Sebastien J-PA Haneuse, James~M Robins, Sarah~E Robertson,
  Ashley~L Buchanan, Elisabeth~A Stuart, and Miguel~A Hern\'an.
\newblock Study designs for extending causal inferences from a randomized trial
  to a target population.
\newblock {\em arXiv preprint arXiv:1905.07764}, 2019.

\end{thebibliography}
\putbib[bibliography]
%%%%%%%%%%%%%%%%%%%%%%%%%%%%%%%%%%%%%%%%%%%%%%%%%%%%%%%%%%%%%%%%%%%%%%%%%%%%%%

\end{bibunit}

\end{document}